\documentclass[a4paper]{article}

\usepackage{cite}
\usepackage[cmex10]{amsmath}
\usepackage{amssymb}
\usepackage{color}
\usepackage[T1]{fontenc}
\usepackage[dvips]{graphicx}

\newcommand{\ch}[1]{{#1}}

\begin{document}
%
\title{\bf Modelling of AC loss in coils made of thin tapes under DC bias current\footnote{Manuscript submitted 16. July 2013. {\color{blue} Published with reference IEEE Trans. Appl. Supercond. (2014) {\bf 24} 4700105, DOI: 10.1109/TASC.2013.2282495.} }}

\author{E. Pardo, \\
Institute of Electrical Engineering, Slovak Academy of Sciences,\\
Dubravska 9, 84104 Bratislava, Slovakia.}

\date{}

\markboth{IEEE Trans. Appl. Supercond. (2014) \bf{24} 4700105, DOI: 10.1109/TASC.2013.2282495}%
{}

\maketitle

\begin{abstract}

Many applications, such as magnets and SMES, are usually charged and discharged under a bias DC current, which may increase the AC loss. For their design, it is necessary to understand and predict the AC loss. This article analises the AC loss in magnet-like coils under DC bias contribution super-imposed to the AC current. The \ch{analyses} is based on a numerical model that takes the interaction between magnetization currents in all turns into account. The studied example is a stack of 32 pancake coils with 200 turns each made of thin tape, such as $Re$BCO coated conductor. We present the current density, the instantaneous power loss, and loss per cycle. We have found that the loss increases with the DC bias current. The instantaneous power loss is the largest in the initial rise of the the AC current. In following cycles, the power loss is higher in the current increase than in the decrease. The loss per cycle is the largest at the end pancakes. In conclusion, the highest cooling power should be supplied to the top and bottom pancakes and during current rise, specially the initial one. The presented model has a high potential to predict the AC loss in magnet-size coils, useful for their design.

{\bf Keywords:} Pancake coil, stack of pancake coils, AC loss, YBCO, coated conductor, numerical modelling

\end{abstract}


\section{Introduction}
\label{s.intro}

Many applications contain windings with a large number of turns, such as magnets, Superconducting Magnetic Energy Storage systems (SMES), transformers and syncronous motors. For their design, it is necessary to understand and predict the AC loss. Too high AC loss difficultates cooling, reduces efficiency and limits the ramp of magnets. Magnets and SMES are usually charged and discharged under a bias DC current in the windings, which may increase the AC loss due to the dynamic magneto-resistance effect \cite{ogasawara76Cry,brandt02PRL}. In addition, windings in the rotor of syncronous motors and generators may experience ramp increase and decrease of current at the start-up or due to torque and power control.

Therefore, it is necessary to understand the main features of the AC loss in magnet-like windings in background DC current. Moreover, numerical modelling tools are required for design and optimization.

In addition, there are limitations in the measurement of instantaneous power loss. Although providing the correct loss per cycle, electrical measurements often read a false power loss signal \cite{acreview,schwerg12IES}. Thermal measurements do not present this problem but their time response is slow. Thus, computations may be the only way to obtain an insight on the instantaneous power loss. 

The present state of the art of numerical computations of coils with many turns is to approximate that the effect of the whole coil in a certain turn is the same as an applied magnetic field. This applied field (``background magnetic field") is computed by assuming that the current density is uniform in the rest of the turns (we name this approach as ``uniform approximation"). Afterwards, the AC loss in the turn of study is estimated by either measurements in a single tape \cite{oomen03SST,kawagoe04IES} or by numerical calculations \cite{tonsho04IES}. The problem of this approximation is that the neighboring turns shield the background magnetic field, in a similar way as in a stack of tapes \cite{pardo03PRB,grilli06PhC}. This shielding effect is very important in windings consisting on stacks of pancake coils made of closely packed thin conductors, such as $Re$BCO coated conductors \cite{pancakenonSC}.

An important step forward for the modelling of single pancake coils has been the continuous approximation, \ch{introduced in \cite{prigozhin11SST} for stacks of tapes, further developped for Finite Element Methods in \cite{zermeno13prp} and applied to circular coils in \cite{pancakenonSC}}. 

The AC loss of a large coil composed of a stack of pancakes under a DC current super-imposed to the alternating excitation has not been published. Existing work is on single pancake coils \cite{hongZ11IES}, racetrack coils under AC applied field \cite{pardo13IES}, \ch{round wires \cite{lahtinen13IES}, and single tapes \cite{tapeDCAC}}.

This article analyses the AC loss in magnet-like coils under DC bias contribution super-imposed to the AC current by means of numerical computations. The studied example is a stack of 32 pancake coils with 200 turns each. The numerical model assumes the sharp $E(J)$ relation of the critical state model ($E$ and $J$ are the electric field and current density, respectively) and takes the interaction of magnetization currents in all turns into account. The computations also take the continuous approximation. For simplicity, we assume constant $J_c$ and normalize all results in order to be independent on $J_c$. The assumptions of critical-state model and constant $J_c$ also allow to separate the phenomena related to flux penetration to those related to magnetic-field dependence of $J_c$ and smooth $E(J)$ relation. \ch{For no DC bias current, we discussed the influence of the magnetic field dependence of $J_c$ on the AC loss in \cite{pardo12SSTb}.}

The article is structured as follows. First, section \ref{s.model} presents details of the studied situation and outlines the numerical model. Section \ref{s.results} discusses the results for the current density, the instantaneous power loss, the loss per cycle and the distribution of loss per cycle between the pancakes composing the coil. Finally, section \ref{s.conclusions} draws the conclusions.



\section{Model}
\label{s.model}

This section first details the studied situation, that is the coil geometry and transport current conditions, and later outlines the numerical model.

\subsection{Studied situation}
\label{s.situation}

The coil \ch{under} study is a stack of 32 pancake coils made of 200 turns each (6400 turns in total), 4 mm wide tape, pancake spacing of 465 $\mu m$, and inner and outer radius of 29.5 and 39.6 mm, respectively. For a 90 $\mu m$ thick tape, such as that from SuperPower \cite{SuperPower}, these radial dimensions corresponds to an inter-turn spacing of 188 $\mu m$. This configuration may be used as a magnet or a SMES. For a critical current of 100 A, the maximum generated magnetic field is 4.7 T. This critical current at the maximum magnetic field in the winding (around 6 T) is achievable between 50 and 65 K for state-of-the-art tapes \cite{selvamanickam12SST}.

The DC and AC currents are taken as follows. The DC current, $I_{DC}$ is set by monotonically increasing the current from the zero-field cool state. Afterwards, the current is further increased to $I_{DC}+I_m$, where $I_m$ is the AC current amplitude. Later the current is periodically decreased and increased to $I_{DC}-I_m$ and $I_{DC}+I_m$, respectively. In the critical-state model, the current density and the loss per cycle do not depend on the waveform, as long as the current increases or decreases monotonically in each half-cycle. However, the instantaneous power loss depends on the particular waveform. In this article, we take the sinusoidal waveform of figure \ref{f.P}.

\subsection{Numerical method}

Our simulations in this article are based on the Minimum Magnetic Energy Variation (MMEV) method. This method assumes the critical-state model in its most general condition, that is $|J|$ can take any value lower or equal to $J_c$ \cite{HacIacinphase,prigozhin97IES}. The main developments to compute superconducting coils are in \cite{pancaketheo}, the implementation of a field dependent $J_c$ in \cite{souc09SST}, interaction with linear magnetic materials in \cite{pancakeFM}, and non-uniform tape properties in \cite{polak12IES}. The present work assumes constant $J_c$ for simplicity. In addition, all results are normalized in a way that they are independent on $J_c$. For this case, all turns present the same critical current, $I_c$, which corresponds to the coil critical current. Note that this is not the case for a magnetic-field dependence of $J_c$, where the coil critical current is that of the weakest turn \cite{coatedIc,zhangM12JAP}.

We calculate the power AC loss by integrating the local power dissipation $E\cdot J$ \cite{acreview} over the volume, after calculating $E$ from the vector potential created by $J$ as detailed in \cite{pancaketheo}. The loss per cycle is the time integration of the power loss in one cycle. Note that integrating in half cycle is not sufficient, due to the assimetry of the loss signal (figure \ref{f.P}). The loss per cycle in this article is for the 3rd cycle after the first current rise.

In order to be able to calculate coils with a large number of turns, we use the continuous approximation \cite{prigozhin11SST}. That is, we approximate the pancake cross-section by a continuous object with the same average critical current density. The current constrain is set by fixing the integral in the axial direction of the current density $J$. \ch{For our case}, it is enough to approximate the real pancake by another one with lower number of turns ($n_{\rm eff}$), \ch{one element per turn,} no separation between turns, the same radial thickness, and effective transport current $I_{\rm eff}=nI/n_{\rm eff}$, where $n$ is the number of turns in one pancake and $I$ is the coil net current \cite{pancakenonSC}.

The model uses the geometrical parameters in section \ref{s.situation}. The numerical method takes a rectangular mesh 50 and 20 elements in the axial and radial directions for each pancake coil, 40 time steps per cycle, 3 cycles and a tolerance of $J$ between 0.25 and 0.036 \% of $J_c$, being the lowest values for the smallest AC fields.


\section{Results and discussion}
\label{s.results}

This section analyses the current density, instantaneous loss, loss per cycle, and loss per cycle per pancake in the coil of study.


\begin{figure}[tbp]
\begin{center}
\includegraphics[width=10cm]{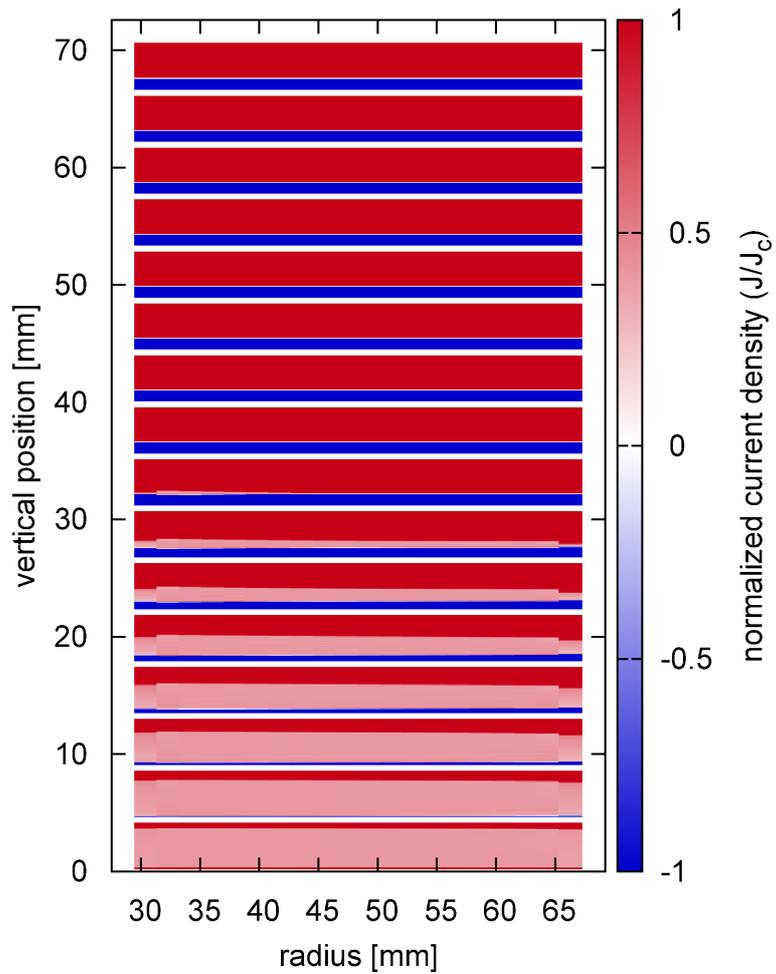}%
\end{center}
\caption{\label{f.Ji} Half of the pancakes of the coil are saturated with critical current density after setting the DC current. Only the upper half of the coil cross-section is shown, although the whole coil is modelled. The situation corresponds to $I_{DC}=0.5I_c$.}
\end{figure}

\begin{figure}[tbp]
\begin{center}
\includegraphics[width=10cm]{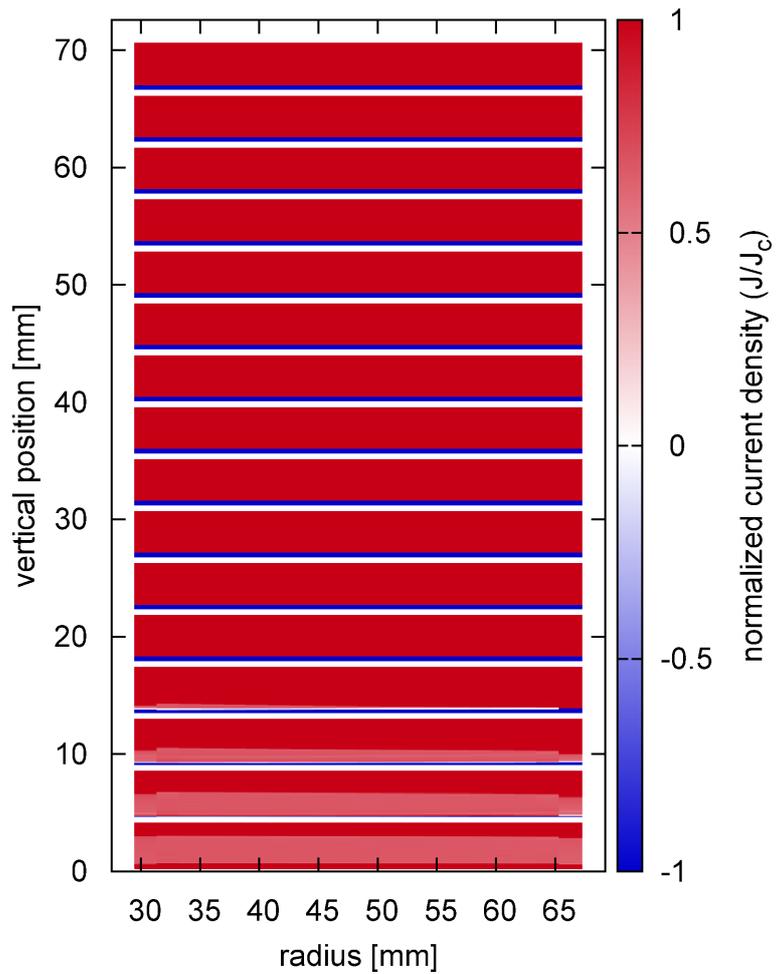}%
\end{center}
\caption{\label{f.Jt} \ch{Current density for the maximum transport current, $I_{DC}+I_m=0.8I_c$, and $I_{DC}=0.5I_c$} for the end of the second increase of current, Fig. \ref{f.P}. Only half of the coil is shown. }
\end{figure}

\begin{figure}[tbp]
\begin{center}
\includegraphics[width=10cm]{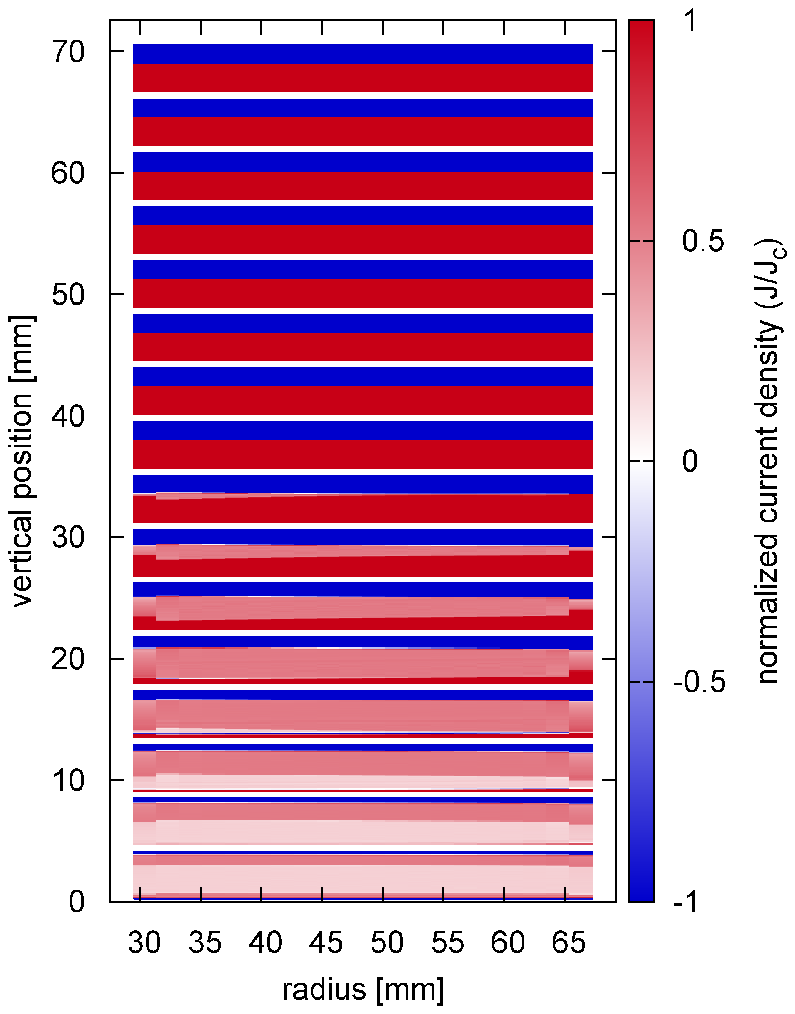}%
\end{center}
\caption{\label{f.Jr} At the minimum of the AC cycle, there appear two sub-critical $J$ in the central pancakes. The presented case is for the end of the first decrease of current, where the transport current is $I_{DC}-I_m=0.2I_c$ and $I_{DC}=0.5I_c$, Fig. \ref{f.P}. Only half of the coil is shown. }
\end{figure}	

The current density after setting the DC current presents strong magnetization currents (see figure \ref{f.Ji} for $I_{DC}=0.5I_c$). These are responsible for the negative $J$, present in all pancakes except the two central ones. The eight top pancakes are saturated. \ch{As expected \cite{pardo12SSTb}, the unsaturated pancakes present sub-critical regions with $J$ roughly uniform and approximately the same in all pancakes. The reason is that this $J$ is proportional to $\partial B_z/\partial r$ (where $r$ and $z$ are the radial and axial coordinates, respectively) \cite{pardo12SSTb} and for our coil this gradient is roughly uniform in the sub-critical region of the unsaturated pancakes. In contrast, for a single pancake this gradient is non-uniform in the $r$ direction, resulting in a non-uniform sub-critical $J$ \cite{prigozhin11SST}.} After increasing the current to $I_{DC}+I_m$, the portion of negative current density is strongly suppressed because of the current constrain, the sub-critical region shrinks, and $J$ there increases (figure \ref{f.Jt} shows the case for $I_{DC}=0.5I_c$ and $I_m=0.3I_c$ \ch{-- actually the program uses $I_m=0.29999I_c$}). Following the AC cycle, the current decreases down to $I_{DC}-I_m$. At that instant, the portion with negative current is larger in the saturated pancakes because the current constrain requires a lower amount of positive $J$ (figure \ref{f.Jr} shows the case for $I_{DC}=0.5I_c$ and $I_m=0.3I_c$). The sub-critical region at central pancakes presents two different $J$ values. The portions with higher sub-critical $J$ correspond to regions with $J=J_c$ at the maximum current, while the lower sub-critical $J$ matches with the sub-critical region at the maximum $J$ (compare figures \ref{f.Jt} and \ref{f.Jr}). The cause of these two sub-critical regions is the following. The current density at the region where has been always in sub-critical state perfectly shields the radial field, perpendicular to the tapes wide side. As a consequence, $J$ is proportional to the gradient of the parallel field, due to Amperes law \cite{roebelcomp}. In the rest of the sub-critical region, the perpendicular field is frozen in the whole half-cycle. Since the change of the parallel field component is proportional to the change of $I$, the decrease in $J$ in the whole sub-critical region is uniform. The values of the sub-critical $J$ are roughly, 0.71$J_c$ at current $I_{DC}+I_m$ and 0.52$J_c$, 0.19$J_c$ for each sub-critical region for current $I_{DC}-I_m$. Figure \ref{f.Jt} is actually for the instant that the current reaches $I_{DC}+I_m$ for the second time, although the current density after the first rise of current is practically identical.

\begin{figure}[tbp]
\begin{center}
\includegraphics[width=10cm]{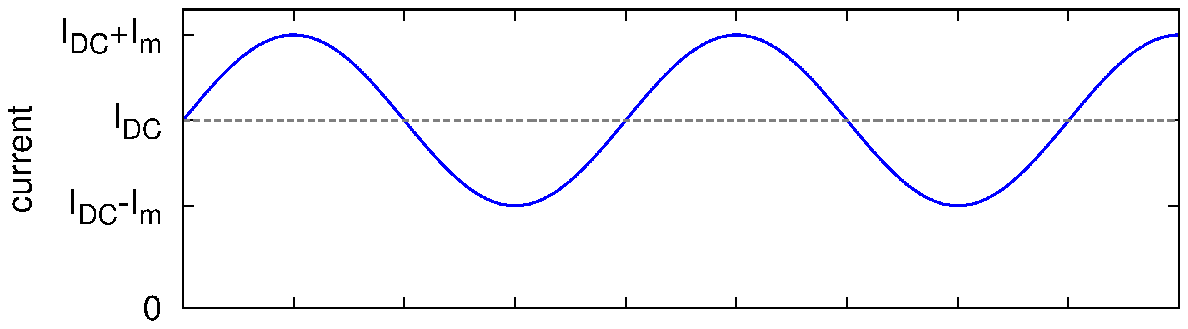}\\%
\includegraphics[width=10cm]{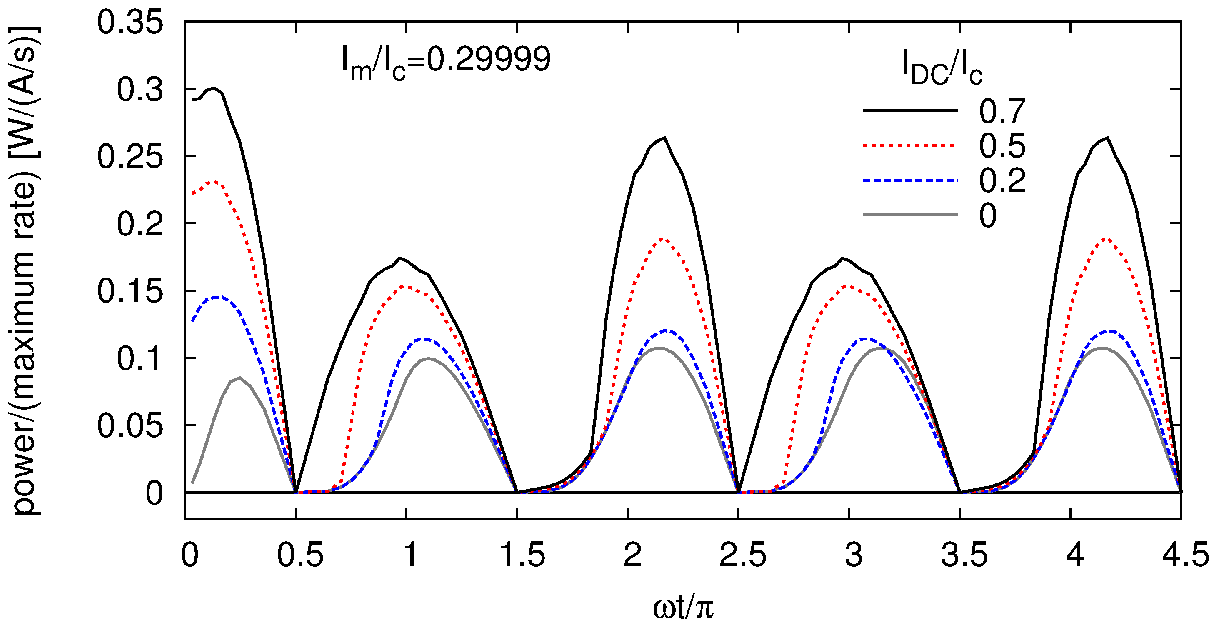}%
\end{center}
\caption{\label{f.P} Top: current waveform after setting the DC current. Bottom: The instantaneous power loss is higher at current rise than at current decrease. The power loss is normalized to the maximum current rate in the AC cycle, appearing at integer $\omega t/\pi$, where $\omega$ is the angular frequency.}
\end{figure}

The power loss increases with the DC current (see figure \ref{f.P} for $I_m=0.3I_c$). The maximum power loss appears for the first rise of current. In addition, the power loss is larger at the following half-cycles where the current increases compared to those where it decreases. The power loss reaches the periodic stationary state after the end of the first decrease of current, although the power loss at the first decrease is practically the same of the second decrease. This is in contrast to single tapes and racetrack coils in AC applied field and DC current \cite{pardo13IES,tapeDCAC}, which require several cycles to reach the stationary state. Coming back to our case, the power loss with a DC current practically corresponds to that of no DC current for a portion of the curve at the beginning of each half-cycle. The part where the curves depart from each other corresponds to the onset of the dynamic magneto-resistance effect.

\begin{figure}[tbp]
\begin{center}
\includegraphics[width=8cm]{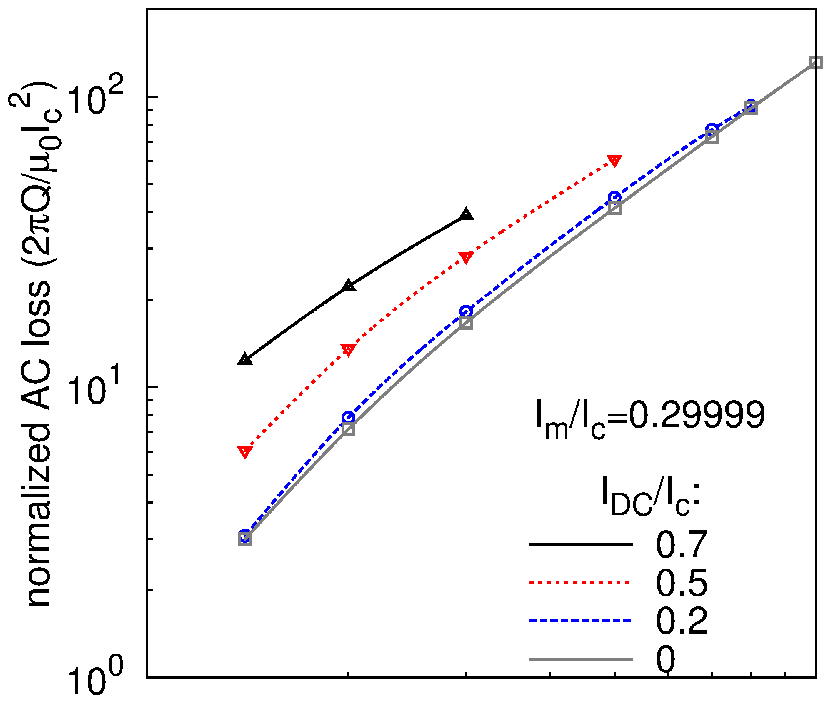}\\%
\includegraphics[width=8cm]{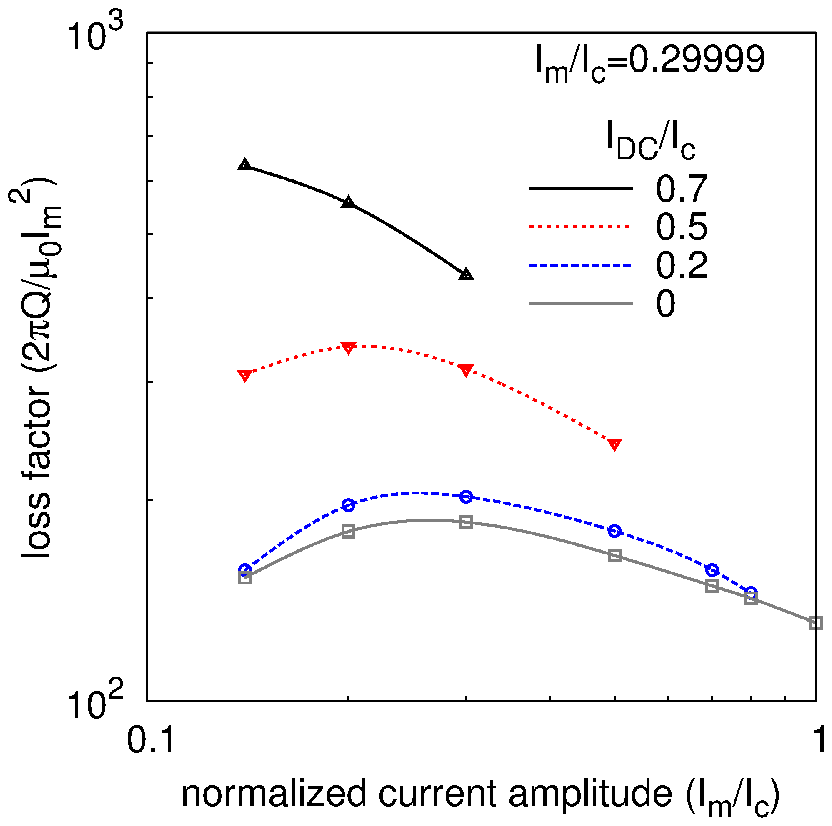}%
\end{center}
\caption{\label{f.QG}
Top: AC loss per cycle and tape length, $Q$, increases with DC current. Bottom: the peak in loss factor evidencies that magnetization loss dominates. In the normalized representation, both quantities are independent on $J_c$.}
\end{figure}

The AC loss per cycle consistently increases with the DC bias current for all the amplitudes (figure \ref{f.QG}a). The relative increase comparing with no bias current becomes larger with decreasing the AC current. The details of the dependence with the current amplitude are more evident for the loss factor, proportional to the loss per cycle divided by $I_m^2$. For single pancakes, this loss factor increases monotonically with the AC current \cite{pancaketheo}. However, for our case the loss factor presents a peak (figure \ref{f.QG}b). This evidents large magnetization currents \cite{HacIacinphase}, in consistence with the results for the current distribution (figures \ref{f.Ji}-\ref{f.Jr}).

\begin{figure}[tbp]
\begin{center}
\includegraphics[width=8cm]{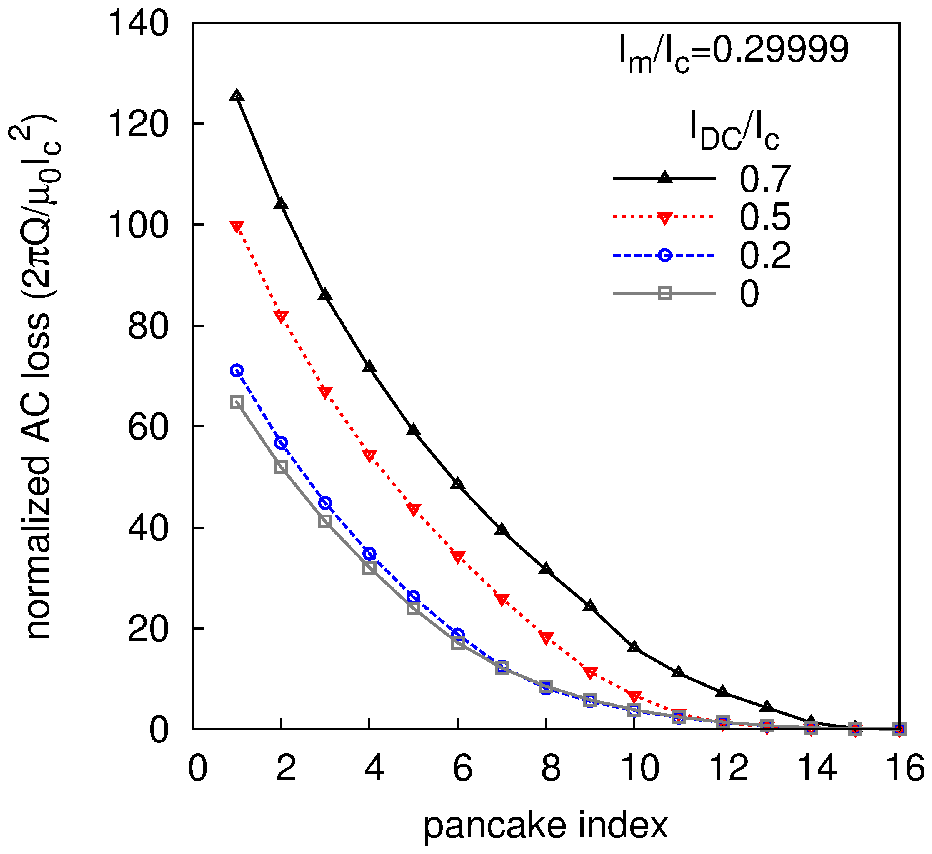}%
\end{center}
\caption{\label{f.Qpan} 
The loss per cycle and tape length, $Q$, is the highest at the end pancakes (pancakes indexed from top to bottom in Fig. \ref{f.Ji}).
}
\end{figure}

The AC loss in each pancake increases with approaching to the coil ends (figure \ref{f.Qpan}). With increasing the DC current, the AC loss not only increases in all pancakes but also the relative contribution from the inner pancakes becomes more important.


\section{Conclusions}
\label{s.conclusions}

Summarizing, this article has presented calculations of the current density, instantaneous power loss and loss per cycle for a magnet-size coil under a DC bias current super-imposed to the AC current. We have studied a particular example of a stack of 32 pancake coils and 200 turns per pancake. For the first time in such magnet-size coils, the model in this article takes the interaction between magnetization currents into account. Actually, the usual approach to assume uniform current density in all turns except the turn where the AC loss is calculated may produce large errors, specially for coils consisting on few pancakes up to considerable large ones (up to around 30 pancakes and thousands of turns in total) \cite{pancakenonSC}. Instead, in order to simplify the problem, we have taken the continuous approximation, which does not introduce additional errors \cite{prigozhin11SST}.

We have found that the top pancakes are easily saturated with magnetization currents. As a consequence, magnetization loss in these pancakes dominates the total loss per cycle. In addition, in parts of the AC cycle, there are two sub-critical current densities in the unsaturated pancakes. The instantaneous power loss is the largest in the initial rise of the the AC current. In following cycles, the power loss is higher in the current increase than in the decrease. For all cases, the loss per cycle is the largest at the end pancakes. \ch{The latter two features are not a consequence of the $J_c(B)$ dependence because they are already present for constant $J_c$. However, a $J_c(B)$ dependence may accentuate these effects.}

In conclusion, the highest cooling power should be supplied to the top and bottom pancakes \ch{(also if the coil is operated well below its critical current)} and during the current rise, specially the initial one. In addition, the presented model has a high potential to predict the AC loss in magnet-size coils, also for a magnetic-field dependence of $J_c$.

\section*{Acknowledgements}

The research leading to these results has received funding from the European Union Seventh Framework Programme [FP7/2007-2013] under grant agreement number NMP-LA-2012-280432. We also acknowledge the support of EURATOM FU-CT-2007-00051 project co-funded by the Slovak Research and Development Agency under contract number DO7RP-0018-12.

\end{document}